\documentclass[twocolumn,showpacs,amsmath,amssymb,pra,superscriptaddress]{revtex4-1}
\usepackage{amssymb}
\usepackage{amsmath}
\usepackage{microtype}
\usepackage {longtable}
\usepackage{bm}
\usepackage{graphicx}
 \usepackage[hyperindex,breaklinks]{hyperref}
\newcommand{\ket}[1]{|#1\rangle}

\newcommand \be{\begin{equation}}
\newcommand \ee{\end{equation}}
\newcommand \bea{\begin{eqnarray}}
\newcommand \eea{\end{eqnarray}}
\newcommand \bse{\begin{subequations}}
\newcommand \ese{\end{subequations}}

\begin{document}
\relpenalty=1
\title{Fast three-qubit Toffoli quantum gate based on the three-body F\"{o}rster resonances in Rydberg atoms}

\author{I.~I.~Beterov}
\email{beterov@isp.nsc.ru}
\affiliation {Rzhanov Institute of Semiconductor Physics SB RAS, 630090 Novosibirsk, Russia}
\affiliation {Novosibirsk State University, 630090 Novosibirsk, Russia}
\affiliation {Novosibirsk State Technical University, 630073 Novosibirsk, Russia}

\author{I.~N.~Ashkarin}
\affiliation {Rzhanov Institute of Semiconductor Physics SB RAS, 630090 Novosibirsk, Russia}
\affiliation {Novosibirsk State  University, 630090 Novosibirsk, Russia}

\author{E.~A.~Yakshina}
\affiliation {Rzhanov Institute of Semiconductor Physics SB RAS, 630090 Novosibirsk, Russia}
\affiliation {Novosibirsk State University, 630090 Novosibirsk, Russia}

\author{D.~B.~Tretyakov}
\affiliation {Rzhanov Institute of Semiconductor Physics SB RAS, 630090 Novosibirsk, Russia}
\affiliation {Novosibirsk State University, 630090 Novosibirsk, Russia}

\author{V.~M.~Entin}
\affiliation {Rzhanov Institute of Semiconductor Physics SB RAS, 630090 Novosibirsk, Russia}
\affiliation {Novosibirsk State University, 630090 Novosibirsk, Russia}

\author{I.~I.~Ryabtsev}
\affiliation {Rzhanov Institute of Semiconductor Physics SB RAS, 630090 Novosibirsk, Russia}
\affiliation {Novosibirsk State University, 630090 Novosibirsk, Russia}

\author{P.~Cheinet}
\affiliation {Laboratoire Aim\'e Cotton, CNRS, Univ. Paris-Sud, ENS-Cachan, Universit\'e Paris-Saclay, 91405 Orsay, France}

\author{P.~Pillet}
\affiliation {Laboratoire Aim\'e Cotton, CNRS, Univ. Paris-Sud, ENS-Cachan, Universit\'e Paris-Saclay, 91405 Orsay, France}

\author{M.~Saffman}
\affiliation {Department of Physics, University of Wisconsin-Madison, Madison, Wisconsin, 53706, USA}

\begin{abstract}

We propose  a scheme of fast three-qubit Toffoli quantum gate for ultracold neutral-atom qubits. The scheme is based on the Stark-tuned  three-body F\"{o}rster resonances, which we have observed in our recent experiment~[D.B.~Tretyakov et al., Phys.~Rev.~Lett. \textbf{119}, 173402 (2017)]. The three-body resonance corresponds to a transition when the three interacting atoms change their states simultaneously, and it occurs at a different dc electric field with respect to the two-body F\"{o}rster resonance. A combined effect of three-body and two-body F\"{o}rster interactions in external electric and magnetic fields near the three-body resonance results in complex coherent behavior of the populations and phases of collective states of a three-atom system. We have found that it is possible to obtain experimental conditions suitable to implement three-qubit Toffoli gate with 98.3\% fidelity and less than 3~$\mu$s duration. 
\end{abstract}
\maketitle

\section{Introduction}

\begin{center}
\begin{figure}[!t]
\center
\includegraphics[width=\columnwidth]{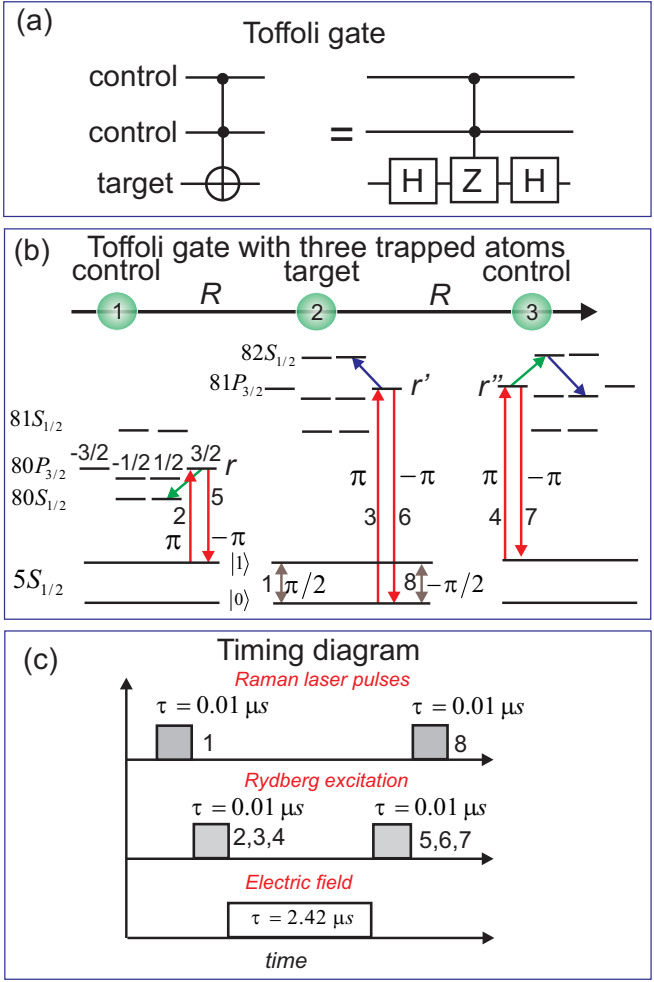}
\vspace{-.5cm}

\caption{
\label{Scheme}
(a) General scheme of the three-qubit Toffoli gate. (b) Scheme of the Toffoli gate  based on three-body Rydberg interactions. Three atoms are located in the individual optical dipole traps along the Z axis directed along the controlling dc electric field. Laser Raman (or microwave) pulses 1 and 8 drive transitions between the logical states $\ket{0}$ and $\ket{1}$ of the target qubit. Laser pulses 2-7 excite and de-excite the chosen Rydberg states of the three atoms. The phase shift due to the two-body interaction of atoms 1 and 3 is $\pi $ if they are both excited into Rydberg states and atom 2 remains in the ground state. Otherwise, all phase shifts are zeroed by the three-body interaction. (c) Timing diagram of the pulses in the proposed three-qubit Toffoli gate.
}

\end{figure}
\end{center}

Three-qubit Toffoli quantum gate is a doubly-controlled NOT gate where the state of the target qubit changes depending on the state of the two control qubits \cite{Toffoli1980}. It can be implemented using a controlled-controlled-Z (CCZ) gate, where the phase of the target qubit is shifted by $\pi$ if both controlling qubits are in state $\ket{1}$, and two Hadamard gates applied to the target qubit before and after the CCZ gate, as shown in Fig.~\ref{Scheme}(a). The Toffoli gate is important for faster implementation of some quantum algorithms, such as the Shor's algorithm~\cite{Shor1999}, and for error correction~\cite{Cory1998, Ottaviani2010,Schindler2011}. This gate in fact can be decomposed as a sequence of six CNOT gates and several single-qubit gates~\cite{Nielsen2011}. Therefore, accumulation of two-qubit errors has a detrimental effect on gate fidelity, and direct single-shot implementation of the Toffoli gate is of great interest. The Toffoli gate has been successfully demonstrated with atomic ions~\cite{Monz2009}, but at a relatively low fidelity of 71\%, and the duration of the gate was as long as 1.5~ms. The Toffoli gate has also been demonstrated with superconducting qubits  at the fidelity of 68.5\% ~\cite{Fedorov2012} and 78\%~\cite{Reed2012}. Another scheme to perform high-fidelity Toffoli gate with superconducting qubits was proposed in Ref.~\cite{Zahedinejad2015}, but it was not implemented experimentally yet. The fidelity of approximately 98\% was achieved only in the experiment with single photons using post-selection techniques~\cite{Lanyon2008}. In the later experiment with linear optics, the fidelity was found to be 83\%~\cite{Micuda2013}.  

In this paper we propose a scheme of fast Toffoli gate for ultracold neutral atoms, which are promising for implementation of a scalable quantum computer~\cite{Brennen1999, Jaksch2000, Saffman2010, Ryabtsev2005, Saffman2016,Ryabtsev2016}. This gate is shown to be much faster than in Refs.~\cite{Monz2009,Fedorov2012,Reed2012} and can provide approximately $3\; \mu s$ duration and 98.3\% fidelity. In contrast to the previously proposed schemes of multibit C\textsubscript{k}NOT gates~\cite{Isenhower2011} and Toffoli gate~\cite{Isenhower2011,Shi2018}, based on Rydberg blockade~\cite{Lukin2001}, we consider  resonant three-body interactions of ultracold Rydberg atoms. Such interactions were first implemented with many Cs atoms in Ref.~\cite{Faoro2015} and then observed by us recently for a few Rb atoms~\cite{Tretyakov2017}.  In our recent complimentary paper~\cite{Ryabtsev2018} we have shown that coherent Rabi-like population oscillations are possible for a three-body F\"{o}rster resonance in spatially localized Rydberg atoms, and they therefore can be used for implementing three-qubit quantum gates.

Although long-range Rydberg interactions couple pairs of the atoms, the three-body interactions in an ensemble consisting of the three atoms result in the simultaneous change of quantum states of all three atoms. They occur at a different dc electric field with respect to the ordinary two-body F\"{o}rster resonances. This can be considered as Borromean energy transfer which features strong three-body interactions while two-body contributions are suppressed~\cite{Faoro2015}. 

The proposed scheme of the Toffoli gate is shown in Fig.~\ref{Scheme}(b).  The three Rb atoms are trapped in the three individual optical dipole traps which are evenly located along the Z axis with distance \textit{R} between neighboring traps. The Z axis is chosen along the controlling dc electric field. In Ref.~\cite{Ryabtsev2018} we revealed that this configuration provides maximum coherence of the population oscillations due to specific selection rules, which close some unwanted interaction channels. The logical states $\ket{0}$ and $\ket{1}$ are hyperfine sublevels of the ground state $5S_{1/2}$. The Toffoli gate can be implemented using 8 laser pulses. To couple the logical states $\ket{0}$ and $\ket{1}$, two-photon Raman laser pulses, which do not populate the intermediate excited state 5\textit{P}~\cite{Brennen1999}, can be used. As an alternative, they can be replaced by microwave pulses with individual addressing. This can be achieved using an intense off-resonant laser acting on a selected qubit to ac Stark shift its energy levels into resonance with the microwave radiation ~\cite{Weitenberg2011,Zhang2006,Xia2015}.

%=========================================================================================================
%=========================================================================================================
%=========================================================================================================
%=========================================================================================================
%=========================================================================================================
%=========================================================================================================

The first Hadamard gate of Fig.~\ref{Scheme}(a) is replaced in Fig.~\ref{Scheme}(b) by the $\pi /2$ rotation around the Y axis, which is implemented by a laser pulse numbered~1 in Fig.~\ref{Scheme}(b). Then, the three laser pulses 2-4 are simultaneously applied to all three qubits. They drive the transitions  $\ket{1}\to\ket{r} $ for the  left control qubit, $\ket{0}\to\ket{r'}$ for the target qubit and $\ket{1}\to\ket{r''} $ for the right control qubit, respectively. Here $\ket{r} =\ket{80P_{3/2} \left({3 /2} \right)}$,  $\ket{r'} =\ket{81P_{3/2} \left({3 /2} \right) } $ and $\ket{r''} =\ket{81P_{3/2} \left({-3 /2} \right)} $. The numbers in the parentheses are the projections of the total momentum $m_J$ on the Z axis. Similarly to the proposal~\cite{Brion2007}, we consider Rydberg excitation in different Rydberg states. Our preliminary analysis has shown that with the selected states it is possible to obtain high fidelity due to long radiative lifetimes, large dipole moments and specific three-body interaction channels. 

We assume that for a selected distance between the atoms Rydberg interaction is sufficiently weak to avoid Rydberg blockade. Depending on the initial state of the atoms, after laser pulses 2-4, the number of the excited Rydberg atoms varies from zero to three. When two or three Rydberg atoms are excited, the interaction shifts the phases of the collective three-body atomic states. These phases can be controlled by external electric and magnetic fields if the interactions are tuned to a three-body F\"{o}rster resonance.  In what follows we show how these phases can be adjusted to implement the Toffoli gate. 

At the final stage, Rydberg atoms are de-excited by laser pulses 5-7. Raman laser or microwave pulse 8 drives the additional $-\pi /2$ rotation of the target qubit around the Y axis, which is equivalent to the second Hadamard gate in Fig.~\ref{Scheme}(a). The timing diagram of all controlling pulses is shown in Fig.~\ref{Scheme}(c) and will be discussed later.

\section{Coherent three-body F\"{o}rster resonances}

Incoherent three-body F\"{o}rster resonances for Rb and Cs Rydberg states have been studied by us previously, both experimentally and theoretically, in the disordered atomic ensembles~\cite{Faoro2015, Tretyakov2017}. In our recent paper~\cite{Ryabtsev2018} we have shown that for the fixed positions of the atoms it is possible to observe coherent three-body F\"{o}rster interactions and Rabi-like population oscillations.

The operator of the dipole-dipole interaction between two neighboring atoms located along the Z axis can be written as~\cite{Walker2008}

\begin{eqnarray}
\label{eq1}
V_{dd} &=&\frac{e^{2} }{4\pi \varepsilon _{0} R^{3} } \left(a\cdot b-3a_{z} b_{z} \right)=\\
&=&-\frac{\sqrt{6} e^{2} }{4\pi \varepsilon _{0} R^{3} } \sum _{q=-1}^{1}C_{1q1-q}^{20} a_{q} b_{-q}.  \nonumber
\end{eqnarray}

Here $\varepsilon _0$ is the dielectric constant, \textit{e} is the electron charge and \textbf{a} and \textbf{b} are the vectorial positions of the two Rydberg electrons. This operator couples only two-atom collective states with $\Delta M=0$, where \textit{M} is the total moment of the collective state. The radial matrix elements of the dipole moment are calculated using a quasiclassical approximation~\cite{Kaulakys1995}.
We now consider collective states of the three Rb Rydberg atoms $\ket{n_{1} l_{1} j_{1} m_{j1} ;n_{2} l_{2} j_{2} m_{j2} ;n_{3} l_{3} j_{3} m_{j3}} $. The F\"{o}rster energy defect is the difference between the energy of an arbitrary final collective state and of the initial collective state. 

The collective states $\ket{80P_{3/2}\left(3/2\right)81P_{3/2}\left(3/2\right)81P_{3/2}\\\left(-3/2\right)}$ and $\ket{80S_{1/2}\left(1/2\right)81P_{3/2}\left(3/2\right)82S_{1/2}\left(-1/2\right)}$ intersect in the electric field of 0.117 V/cm which corresponds to the \textbf{two-body} F\"{o}rster resonance  $\ket{80P_{3/2}\left(3/2\right)81P_{3/2}\left(-3/2\right)}\to \ket{80S_{1/2}\left(1/2\right)82S_{1/2}\left(-1/2\right)}$. In a zero electric field the F\"{o}rster energy defect for this resonance is around 110~MHz. 

%=========================================================================================================
%=========================================================================================================
%=========================================================================================================
\begin{center}
\begin{figure}[!t]
\center
\includegraphics[width=\columnwidth]{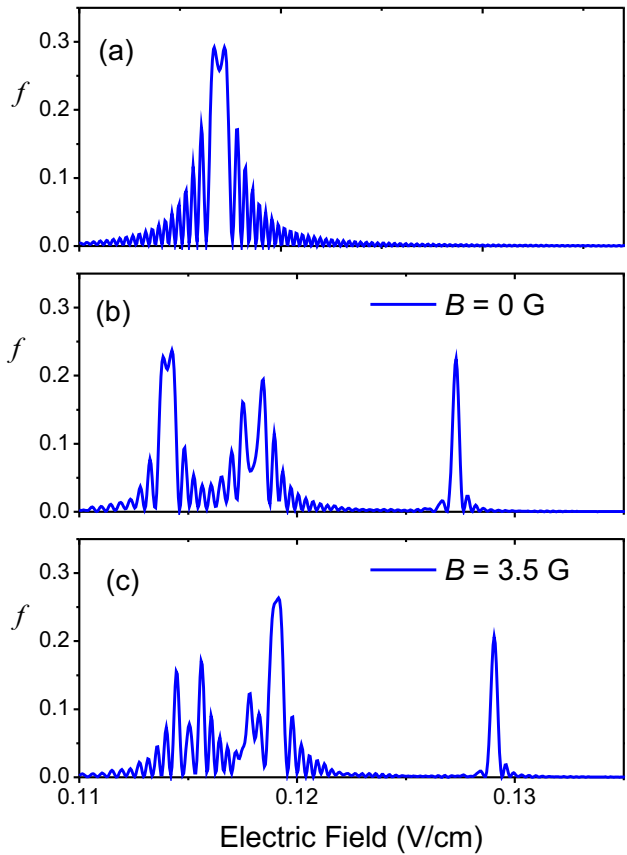}
\vspace{-.5cm}
\caption{
\label{Stark}
Numerically calculated dependence of the fraction $f$ of atoms in the final $\ket{80S_{1/2}}$ state on the dc electric field: (a) for two atoms initially prepared in the collective state $\ket{80P_{3/2}(3/2) 81P_{3/2} (-3/2)}$ at the interatomic distance $R=25\; \mu $m and interaction time $\tau =1.8\; \mu $s; (b) for the three-atom ensemble initially prepared in the state $\ket{80P_{3/2}(3/2) 81P_{3/2} (3/2) 81P_{3/2} (-3/2)}$ at the interatomic distance $R=12.5\; \mu $m and interaction time $\tau =1.8\; \mu $s in the external magnetic field \textit{B}=0~G; (c) for the three-atom ensemble initially prepared in the state $\ket{80P_{3/2}(3/2) 81P_{3/2} (3/2) 81P_{3/2} (-3/2)}$ at the interatomic distance $R=12.5\; \mu $m and interaction time $\tau =1.8\; \mu $s in the external magnetic field \textit{B}=3.5~G. 
}
\end{figure}
\end{center}
%=========================================================================================================
%=========================================================================================================
%=========================================================================================================

The two-body F\"{o}rster resonance is observed as increase of the fraction $f$ of atoms in the final state $\ket{80S_{1/2}}$ if two atoms are initially prepared in a collective state $\ket{80P_{3/2}\left(3/2\right)81P_{3/2}\left(-3/2\right)}$. The numerically calculated  dependence of $f$ on the dc electric field for two atoms at interatomic distance $25\,\mu$m and interaction time $1.8\,\mu$s is shown in Fig.~\ref{Stark}(a). The resonance is observed in the electric field of 0.117~V/cm, which corresponds to the position of the two-body F\"{o}rster resonance. This resonance is insensitive to the magnetic field due to the compensation of the Zeeman shifts for two-atom $\ket{PP}$ and $\ket{SS}$ collective states with $M=0$.

If we place a third atom in the  state $\ket{81P_{3/2} \left(3/2\right)}$ between  the two atoms, which are prepared in states $\ket{80P_{3/2} \left(3/2\right)}$ and $\ket{81P_{3/2} \left(-3/2\right)}$, respectively, additional three-body resonances will arise. The collective states  $\ket{80P_{3/2}\left(3/2\right)81P_{3/2}\left(3/2\right)81P_{3/2}\left(-3/2\right)}$ and $\ket{80S_{1/2}\left(1/2\right)82S_{1/2}\left(1/2\right)81P_{3/2}\left(1/2\right)}$  intersect in the electric field of 0.123 V/cm. This is an example of a \textbf{three-body} F\"{o}rster resonance. The three-body interaction couples collective three-atom states $\ket{80P_{3/2}\left(3/2\right)81P_{3/2}\left(3/2\right)81P_{3/2}\left(-3/2\right)}$ and $\ket{80S_{1/2}\left(1/2\right)82S_{1/2}\left(1/2\right)81P_{3/2}\left(1/2\right)}$  through the intermediate state $\ket{80S_{1/2}\left(1/2\right)81P_{3/2}\left(3/2\right)82S_{1/2}\left(-1/2\right)}$, as can be seen in Fig.~\ref{Scheme}(b). We note that other three-body interaction channels through different intermediate states also exist and may contribute to the population transfer and phase shifts~\cite{Ryabtsev2018}. 

The spatial configuration which we selected is advantageous for the observation of three-body interactions due to suppression of the resonant two-body interaction between pairs of atoms despite the large interaction energies used. Due to the selection rule $\Delta M=0$, collective state of atoms 1 and 2 $\ket{80P_{3/2}\left(3/2\right)81P_{3/2}\left(3/2\right)}$  is not coupled to state $\ket{80S_{1/2}\left(1/2\right)82S_{1/2}\left(1/2\right)}$. Therefore, the two-body interaction between atoms 1 and 2 is forbidden for these particular states. The two-body interaction between atoms 2 and 3, which are both excited into $\ket{81P_{3/2}} $ state, is suppressed due to the large (157~MHz) F\"{o}rster energy defect for the two-body F\"{o}rster interaction channel $\ket{81P_{3/2}\left(3/2\right)81P_{3/2}\left(-3/2\right)}\to \ket{81S_{1/2}\left(1/2\right)82S_{1/2}\left(-1/2\right)}$  in zero electric field, which increases further to 283~MHz when resonant electric field of 0.123~V/cm is applied. However, we should additionally take into account the always-resonant exchange interaction appearing due to excitation hopping between $S$ and $P$ Rydberg atoms~\cite{Faoro2015}. It drives the unwanted transition $\ket{81P_{3/2} \left(3/2 \right) 81P_{3/2}  \left(-3/2\right)} \to \ket{ 81P_{3/2} \left(-{3 /2} \right) 81P_{3/2} \left(3 /2 \right)}$ through several intermediate states.

Finally, the interaction between atoms 1 and 3, which are prepared in states $\ket{80P_{3/2}}$ and $\ket{81P_{3/2}}$ is enhanced by the two-body F\"{o}rster resonance $\ket{80P_{3/2}\left(3/2\right)81P_{3/2}\left(-3/2\right)}\to \ket{80S_{1/2}\left(1/2\right)82S_{1/2}\left(-1/2\right)}$. At the same time, it is reduced due to the twice larger distance between these atoms compared to the distance between the neighboring atoms. 

 We have numerically  calculated the dependence of the fraction $f$ of atoms in the final state $\ket{80S_{1/2} }$ on the dc electric field for the ensemble of three Rydberg atoms, initially prepared in the collective state $\ket{80P_{3/2}\left(3/2\right)81P_{3/2}\left(3/2\right)81P_{3/2}\left(-3/2\right)}$ at interatomic distance of $12.5\, \mu$m and interaction time $1.8\,\mu$s. In our simulations, developed originally in Refs.~\cite{Tretyakov2017,Ryabtsev2018}, we have taken into account all magnetic sublevels of the  $\ket{80S_{1/2}}$, $\ket{81S_{1/2}}$, $\ket{82S_{1/2}}$, $\ket{80P_{1/2}}$, $\ket{80P_{3/2}}$, $\ket{81P_{1/2}}$, and $\ket{81P_{3/2}}$ Rb Rydberg states and considered only collective three-atom states with $M=3 /2$, since the total moment of the collective state does not change for the spatial configuration which we have chosen. To reduce the complexity of the calculations, we neglected far-detuned collective states with the F\"{o}rster energy defect larger than 1~GHz. The influence of neighboring \textit{D} states is negligible due to large F\"{o}rster defects of order of 10~GHz.  

The Stark shift of the Rydberg states in an external electric field was taken into account using  polarizabilities of the Rydberg states, numerically calculated for a single atom in an external dc electric field~\cite{Zimmerman1979}. We also introduced a Zeeman splitting of the Rydberg states in an external magnetic field in order to lift the degeneracy of magnetic sublevels.

We solved the Schr\"{o}dinger equation for the probability amplitudes of the 165 collective states taking into account Rydberg lifetimes~\cite{Beterov2009}. For simplicity, we consider an open system and neglect the return of the population from Rydberg to the ground state due to spontaneous decay. The time dynamics of the probability amplitudes in the open system with finite lifetimes of the Rydberg state \textit{i} can be described in the time-dependent Schr\"{o}dinger equation by adding $-{\gamma_{i} /2} $ to the right-hand side of each equation for the probability amplitude~\cite{Shore2008}, where $\gamma_{i}$ is a decay rate of the Rydberg state at the 300~K ambient temperature~\cite{Beterov2009}. The decay rates of the collective three-atom states are calculated by summing up the decay rates of individual atoms.

The calculated dependence of the fraction $f$ of atoms  in the final state $\ket{80S_{1/2}\left(1/2\right)} $ on the controlling dc electric field for three interacting atoms in the magnetic field is shown in Fig.~\ref{Stark}(b) for \textit{B}=0~G  and in Fig.~\ref{Stark}(c) for \textit{B}=3.5~G. The direction of the magnetic field is opposite to the direction of the electric field in order to shift the three-body resonances to the right-hand side. Narrow three-body peaks are clearly observed in Figs.~\ref{Stark}(b) and \ref{Stark}(c). The coherent three-body resonance is split due to always-resonant exchange interactions~\cite{Ryabtsev2018}. Therefore the positions of the three-body resonances are different from 0.123~V/cm if the interaction energy is sufficiently large~\cite{Ryabtsev2018}. The shape of the two-body resonance is modified due to the three-body incoherent population transfer to state $\ket{80S_{1/2}}$. The incoherent transfer occurs when the intermediate state (for example, $\ket{80S_{1/2} \left(1/2\right) 81P_{3/2} \left(3/2 \right) 82S_{1/2} \left(-1/2 \right)}$ is populated. In contrast to  Fig.~\ref{Stark}(a), the positions of the resonances  in Figs.~\ref{Stark}(b) and \ref{Stark}(c) are sensitive to the external magnetic field.

\begin{center}
\begin{figure}[!t]
\center
\includegraphics[width=\columnwidth]{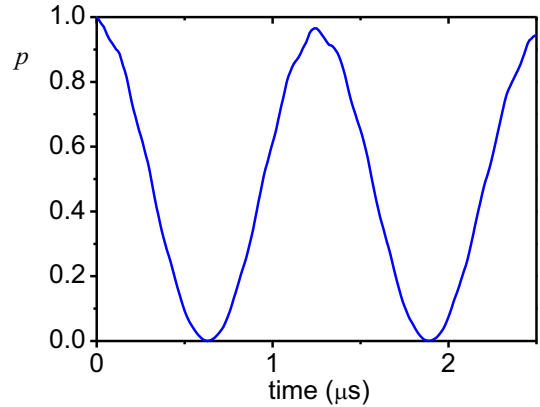}
\vspace{-.5cm}
\caption{
\label{Rabi}
Time dependence of the population $p$ of the initial state $\ket{80P_{3/2} \left(3/2\right) 81P_{3/2} \left(3.2 \right)81P_{3/2}\left(-3/2 \right)}$ for the interatomic distance $R=12.5\; \mu$m and  the external magnetic field \textit{B}=3.5~G and the electric field \textit{E}=0.11905~V/cm.
}
\end{figure}
\end{center}

If we tune the electric field to the resonant value \textit{E}=0.11905~V/cm in magnetic field \textit{B}=3.5~G, coherent Rabi-like population oscillations are observed for the probability $p$ to find the ensemble back in the initially prepared state, as shown in Fig.~\ref{Rabi}. This is a remarkable feature of the three-body F\"{o}rster resonance, which has not been demonstrated experimentally yet~\cite{Ryabtsev2018}. The decrease in the contrast of the oscillations as time increases is mostly caused by finite lifetimes of the Rydberg states, although the leakage of the population to other collective three-atom Rydberg states can also be important.

%=========================================================================================================
%=========================================================================================================
%=========================================================================================================
%=========================================================================================================
%=========================================================================================================
%=========================================================================================================

The coherent evolution of the probability $p$ can be observed in various sets of experimental parameters~\cite{Ryabtsev2018}. In principle, a controlled-phase gate can be based on coherent Rabi-like oscillations at exact resonance, which should lead to a $\pi$ phase shift of the collective state after the end of the $2\pi$ pulse. This is similar to  two-qubit controlled phase gates, considered in our previous works~\cite{Saffman2010, Ryabtsev2005, Beterov2018}. However, off-resonant two-body F\"{o}rster interactions result in a complex behavior of the phase of the collective state. Therefore, below we consider an alternative scheme. The $\pi$ phase shift results from the off-resonant two-body Rydberg interaction. The three-body F\"{o}rster resonance is used only to tune the phase shift to zero, when all three atoms are excited into Rydberg states.

\section{Three-qubit Toffoli gate}

\begin{center}
\begin{figure}[!t]
\center
\includegraphics[width=\columnwidth]{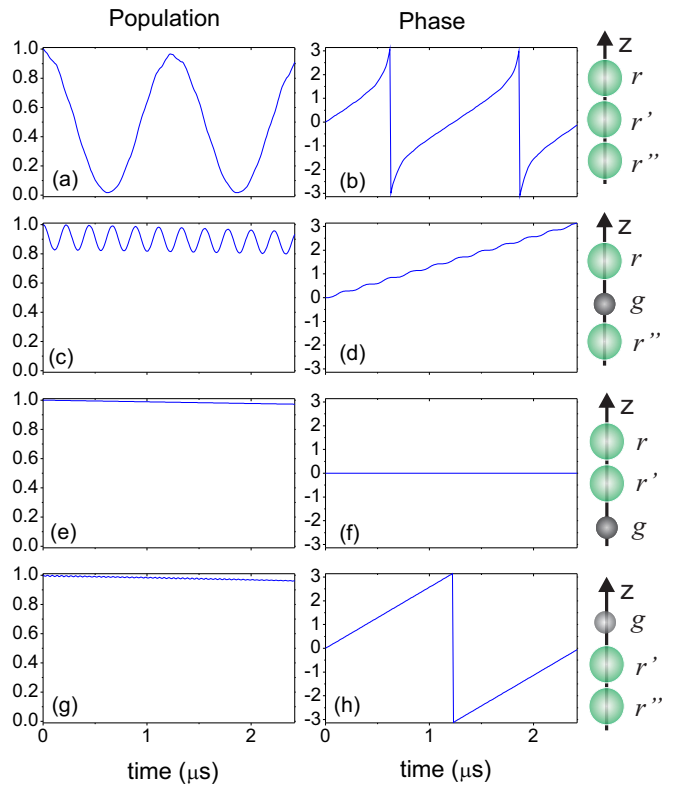}
\vspace{-.5cm}

\caption{
\label{Toffoli}
Numerically calculated time dependences of the population and phase of the collective states of three interacting atoms $\ket{rr'r''} $[(a),(b)]; $\ket{rgr''} $[(c),(d)]; $\ket{rr'g } $[(e),(f)]; $\ket{ gr'r''} $[(g),(h)]. The interatomic distance is $R=12.5\; \mu \text{m}$, the electric field is \textit{E}=0.11912~V/cm and the magnetic field is \textit{B}=3.5~G.
}

\end{figure}
\end{center}

For the implementation of the Toffoli gate it is necessary to find the conditions where the three-body and two-body interactions result in the correct phase shifts of the initially excited collective states. An example is shown in Fig.~\ref{Toffoli}. We numerically calculated the time dependences of the population and phase of the initially excited collective two-atom or three-atom Rydberg states in the electric field \textit{E}=0.11912~V/cm and magnetic field  \textit{B}=3.5~G for the interatomic distance $R=12.5\; \mu $m. The electric field is very close to exact resonance, and it is supposed to be set with high precision, as the three-body resonances are extremely narrow. Small variation of the electric field from the exact resonance value \textit{E}=0.11905~V/cm is used to reduce the phase error. If collective three-atom Rydberg state $\ket{rr'r''}$ is initially excited, we observe almost-resonant Rabi-like oscillations. After time $\tau =2.42\; \mu $s the system returns to the initial state with almost zero phase shift, as shown in Figs.~\ref{Toffoli}(a) and \ref{Toffoli}(b), respectively.

If the ensemble is initially excited into state $\ket{rgr''}$ [here \textit{g} is the ground state which can be either $\ket{0} $ or $\ket{1}$ logic state, depending on the location of the atom in Fig.~\ref{Scheme}(b)], the off-resonant two-body F\"{o}rster interaction $\ket{80P_{3/2} 81P_{3/2}} \to \ket{80S_{1/2}82S_{1/2}}$ shifts the phase of the initially excited state by $\pi $, as shown in Figs.~\ref{Toffoli}(c) and \ref{Toffoli}(d). This phase shift is sensitive to the electric field which acts directly on the F\"{o}rster defect. It corresponds to the controlled phase shift when all three atoms are in state $\ket{1}$ prior to the Rydberg excitation in Fig.~\ref{Scheme}(b). The population of the initial state is reduced by approximately 10\% due to finite Rydberg lifetimes and leakage of the population to other collective states by Rydberg interaction. These are found to be the major sources of the gate error.

%=========================================================================================================
%=========================================================================================================
%=========================================================================================================
%=========================================================================================================
%=========================================================================================================
%=========================================================================================================

If the ensemble is initially excited into  state $\ket{ rr'g } $ [Figs.~\ref{Toffoli}(e) and ~\ref{Toffoli}(f)], there is no interaction due to the selection rule $\Delta M=0$, and the probability to find the ensemble in the initial state reduces only due to finite Rydberg lifetimes.

If the ensemble is initially excited into state $\ket{gr'r''} $ [Figs.~\ref{Toffoli}(g) and ~\ref{Toffoli}(h)], the always-resonant two-body exchange interaction $\ket{81P_{3/2} \left(3/2\right) 81P_{3/2} \left(-3/2 \right)} \to \ket{81P_{3/2} \left(-3/2 \right) 81P_{3/2}  \left(3/2\right) }$ occurs in a multistep manner via intermediate $S$ states, including $\ket{81S\,82S}$ collective state. This leads to the phase shift of the initially excited state, which is compensated for to zero in Fig.~\ref{Toffoli}(h). This phase shift is found to be weakly sensitive to the electric field.

Finally, when only one atom in the ensemble is temporarily excited into the Rydberg state $\ket{r}$, $\ket{r'}$ or $\ket{r''}$, the $\pi$ and $-\pi$ pulses, shown in Fig.~\ref{Scheme}, will return the system into the initial state with zero phase shift. However, temporary Rydberg excitation will result in population loss due to the finite lifetimes of Rydberg states. The trivial case is when no Rydberg atoms are excited. The pulses 2-7 will have no effect in this case.

We applied the following optimization procedure to find the experimental conditions used in Fig.~\ref{Toffoli}: (I) For the selected interatomic distance we find the value of the magnetic field when the two-body and three-body F\"{o}rster resonances weakly overlap, as shown in Figs.~\ref{Stark}(b),(c). This overlapping should be weak enough to avoid distortion of coherent Rabi-like oscillations, shown in Fig.~\ref{Toffoli}(a). At the same time, two-body interactions must be sufficiently large to produce the $\pi$ phase shift in Fig.~\ref{Toffoli}(d). during the interaction time, which is substantially smaller than Rydberg lifetime. (II) We select the value of the electric field near the three-body resonance and find the time interval that compensates for the phase shift due to exchange interaction, as shown in Fig.~\ref{Toffoli}(h). (III) We adjust the value of the electric field to obtain the phase shift close to $\pi $ for the off-resonant two-body F\"{o}rster interaction [Fig.~\ref{Toffoli}(d)]. (IV) We adjust the value of the magnetic field to tune the frequency of three-body Rabi-like oscillations and to find the conditions when the system returns back to the initial state with zero phase shift, as shown in Figs.~\ref{Toffoli}(a) and ~\ref{Toffoli}(b). 

If the last step is not successful, a different interatomic distance should be selected. A compromise has to be found as at larger interatomic distances longer time intervals are required for phase accumulation. Therefore, the fidelity decreases due to finite Rydberg lifetimes. At the same time, at smaller interatomic distances both overlapping of the three-body and two-body resonances and exchange interactions become larger, and it is difficult to find suitable experimental conditions. 

\begin{center}
\begin{figure}[!t]
\center
\includegraphics[width=\columnwidth]{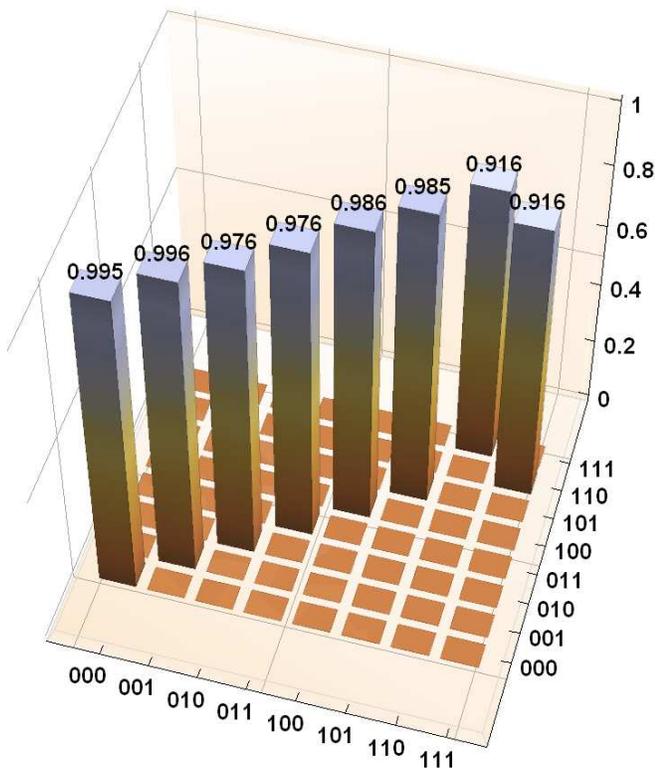}
\vspace{-.5cm}

\caption{
\label{Fidelity}
Numerically simulated truth table for our scheme of the Toffoli gate.
}
\end{figure}
\end{center}

We have numerically calculated the truth table for our scheme of the Toffoli gate (Fig.~\ref{Fidelity}). The timing diagram of the laser pulses is shown in Fig.~\ref{Scheme}(c). In the numerical model, pulses 1-8 have the duration of 10~ns and are applied in a zero electric field. After the laser excitation ends, the electric field \textit{E}=0.11912~V/cm is switched on. The magnetic field \textit{B}=3.5~G  is always present, since it cannot be rapidly switched in experiments. Short laser pulses with high intensity are required to reduce the effect of Rydberg blockade by increasing the Rabi frequencies, which leads to a blockade breakdown. The phase shift of the Rydberg energy levels due to the non-zero electric field is compensated for by adjusting the phases of laser pulses~5-7. 
%=========================================================================================================
%=========================================================================================================
%=========================================================================================================
%=========================================================================================================
%=========================================================================================================
%=========================================================================================================

To estimate the gate fidelity, we have used the method proposed in Ref.~\cite{Bowdrey2002}. We considered 6 single-qubit configuration states $\ket{0}$, $\ket{1}$,$\left(\ket{0}+\ket{1}\right)/\sqrt{2}$, $\left(\ket{0} -\ket{1}\right)/\sqrt{2}$, $\left(\ket{0} +i\ket{1}\right)/\sqrt{2}$ and $ \left(\ket{0} -i\ket{1} \right)/\sqrt{2}$. We formed a set of three-qubit states as all $6^3=216$ combinations of three single-qubit basis states. We simulated the density matrices $\rho_{sim}$ of all final states after Toffoli gate was applied to each initial state. Then we calculated the fidelity of each final state comparing to the etalon state $\rho_{et}$, which is the final state of the ensemble after the ideal Toffoli gate is performed, as follows~\cite{Nielsen2011}:

\begin{equation}
\label{eq2}
F=\text{Tr}\sqrt{\sqrt{\rho _{et} }\rho _{sim} \sqrt{ \rho _{et}}}.  
\end{equation}

After averaging over all 216 states the simulated gate fidelity reaches 98.3\%. It is limited mainly by decay of Rydberg states, undesirable population transfer to different Rydberg states due to off-resonant Rydberg interactions and finite efficiency of phase compensation at three-body F\"{o}rster resonance. Although this value is below the threshold value of 99.99\% for error-tolerant quantum computation, it is substantially better than the value reported so far for ions~\cite{Monz2009}, and the Rydberg Toffoli gate can be much faster that for ions, as well. We believe that our scheme can also be useful for many-body quantum simulations with Rydberg atoms in optical lattices~\cite{Liu2008, Peng2009, Hammer2013,Jachymski2016, You2016}. Additional decrease of gate fidelity can result from dephasing between the laser pulses due atomic motion, which leads first to additional laser phase uncertainty and second to fluctuations of the interatomic distance, which affect the energy of Rydberg interaction. Coherent Rydberg excitation by short laser pulses with large Rabi frequencies which are necessary to avoid Rydberg blockade is also a challenging task. However, such technical difficulties are present for most of the schemes of quantum computing based on Rydberg atoms~\cite{Saffman2016}.

\section{Conclusion}

We proposed a scheme to implement a three-qubit Toffoli gate based on resonant Borromean three-body interactions in the ensemble of Rb Rydberg atoms trapped in three individual optical dipole traps. The collective phase shifts induced by Rydberg interactions are controlled by external electric and magnetic fields. We have shown that it is possible to reach a fidelity exceeding 98\% for a short gate duration of less than \textbf{$3\; \mu \text{s}$}. No high-fidelity multi-qubit gates with ultracold atoms have been demonstrated experimentally yet. The scheme we propose relies on the relatively weak Rydberg interactions. This reduces the effect of the complex structure of Rydberg energy levels on gate fidelity, which appears to be the major source of gate error if the Rydberg interactions are strong~\cite{Derevianko2015}.

This work was supported by Russian Science Foundation Grant No~16-12-0083 in the part of simulation of the Toffoli gate and Grant No 18-12-00313 in the part of the simulation of three-body interactions, RFBR Grant No~17-02-00987 in the part of simulation of the off-resonant Rydberg interactions and Grant No~16-02-00383 in part of simulation of coherent three-body F\"{o}rster resonances, and Novosibirsk State University, the public Grant CYRAQS from Labex PALM (ANR-10-LABX-0039) and the EU H2020 FET Proactive project RySQ (Grant No. 640378). MS was supported by the ARL-CDQI program through cooperative agreement~W911NF-15-2-0061 and NSF award~1720220.

%\bibliography{JCbib}{}

%
\end{document}